\documentclass[aps,twocolumn,floatfix]{revtex4}

\usepackage{graphicx}
\usepackage{dcolumn}
\usepackage{bm}
\usepackage{amsmath}
\begin{document}

\newcommand{\bk}{{\bf k}}
\newcommand{\bp}{{\bf p}}
\newcommand{\bv}{{\bf v}}
\newcommand{\bq}{{\bf q}}
\newcommand{\tbq}{\tilde{\bf q}}
\newcommand{\tq}{\tilde{q}}
\newcommand{\bQ}{{\bf Q}}
\newcommand{\br}{{\bf r}}
\newcommand{\bR}{{\bf R}}
\newcommand{\bB}{{\bf B}}
\newcommand{\bA}{{\bf A}}
\newcommand{\bE}{{\bf E}}
\newcommand{\bj}{{\bf j}}
\newcommand{\bK}{{\bf K}}
\newcommand{\cS}{{\cal S}}
\newcommand{\vd}{{v_\Delta}}
\newcommand{\tr}{{\rm Tr}}
\newcommand{\kslash}{\not\!k}
\newcommand{\qslash}{\not\!q}
\newcommand{\pslash}{\not\!p}
\newcommand{\rslash}{\not\!r}
\newcommand{\bs}{{\bar\sigma}}

\title{Anyons in a weakly
interacting system}

\author{C. Weeks, G. Rosenberg, B. Seradjeh and M. Franz}
\affiliation{Department of Physics and Astronomy,
University of British Columbia, Vancouver, BC, Canada V6T 1Z1}
\date{\today}

\begin{abstract}
In quantum theory, indistinguishable particles in three-dimensional space 
behave in only two distinct ways. Upon interchange, their wavefunction maps 
either to 
itself if they are bosons, or to minus itself if they are fermions. In two 
dimensions a more exotic possibility arises: upon exchange of two particles 
called ``anyons'' the wave function acquires phase $e^{i\theta}\neq\pm 1$.
Such fractional exchange
statistics are normally regarded as the hallmark of strong correlations.
Here we describe 
a theoretical proposal for a system whose excitations are anyons with the 
exchange phase $\theta=\pi/4$ and charge $-e/2$, but, remarkably, can be built 
by filling a set of single-particle states of
essentially noninteracting electrons. The system consists of an artificially 
structured type-II superconducting film adjacent to a 2D electron gas 
in the {\em integer} quantum Hall regime with unit filling fraction. 
The proposal rests on the observation that a vacancy in an otherwise periodic
vortex lattice in the superconductor creates a bound state in the 2DEG with 
total charge $-e/2$. A composite of this fractionally charged hole and the 
missing flux due to the vacancy behaves as an anyon.
The proposed setup allows for manipulation of these anyons 
and could prove useful in various schemes for fault-tolerant 
topological quantum computation.

\end{abstract}
\maketitle

\section{Introduction}

Anyons and fractional charges appear in a variety of theoretical models 
involving electron or spin degrees of freedom in 2D 
\cite{laughlin1,arovas1,wen1,kitaev1,freedman1,kitaev2}. 
In all known cases, interactions between
the fundamental degrees of freedom are of paramount importance for the 
emergence of the exotic excitations. For example, it is understood that the Coulomb
interaction between electrons is crucial to stabilize the incompressible 
ground state in the fractional quantum Hall (FQH) liquids that support
fractionalized excitations.
One can say that such systems are {\em strongly correlated} in the sense
that their many-body wavefunctions cannot be constructed from the 
single-particle states of the constituent fundamental degrees of freedom.
 
The question thus arises whether anyons and fractionalization are inextricably
connected with the phenomenon of strong correlations as defined above. There
are compelling reasons to believe that this is in fact not so. In one dimension, 
Su, Schrieffer and Heeger \cite{su} showed that a {\em half of an electron} 
can be bound to a domain wall in {\em trans}-polyacetylene modeled by a 
dimerized tight binding chain. Interactions play no role in this construction 
and the many-body wavefunction
indeed can be built as a Slater determinant of single-particle electron 
states. In this case fractionalization is a many-body effect in that it is
a cooperative phenomenon of a large number of electrons but it arises from the 
geometry of the system rather than strong correlations.

In three dimensions, Jackiw and Rebbi \cite{jackiv1} 
argued that it is consistent to
assign a {\em half-integer} fermion number to a monopole in the Yang-Mills 
gauge field coupled to relativistic Dirac fermions. In essence, this 
implies that a half-fermion is permanently bound to such a monopole. 
Again, interactions between fermions play no role in this construction and 
the half-fermion emerges from a many-body wavefunction that is a Slater
determinant of single-particle states. 
It has been pointed out
very recently \cite{chamon1} that a two-dimensional version of the
physics underlying the above effect 
could be observable under certain conditions in graphene.

\begin{figure}
\includegraphics[width = 8.0cm]{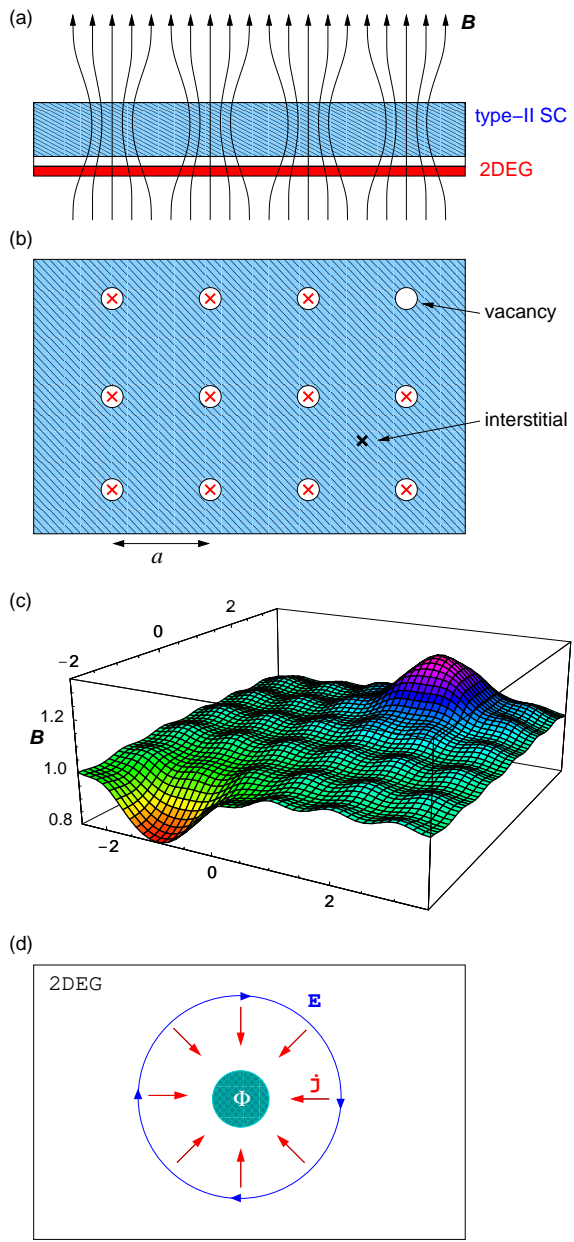}
\caption{(a) The schematic of the proposed device. Side view of
the superconductor-2DEG heterostructure in an applied magnetic field $\bB$. 
(b) Top view. Open circles represent pinning sites
in the superconductor while crosses denote vortices. Throughout this paper 
we consider, for simplicity, the square vortex lattice. All of our conclusions, 
however, remain valid for any arbitrary periodic Bravais lattice, including
the most natural triangular case. 
(c) The magnetic field produced by a periodic square lattice of vortices
with one vacancy (lower left corner) and one interstitial (upper right corner).
The field distribution is calculated from a simple London model \cite{tinkham}
with the penetration depth $\lambda=a$ and the vortex core cutoff 
$\xi=0.05\lambda$. (d) Thought experiment: creating an interstitial by 
adiabatically threading flux through the system.
}
\label{fig1}
\end{figure}

An appealing feature of weakly interacting systems is that their properties 
can be often understood from essentially exact calculations, which is rarely 
possible for strongly correlated systems in dimension greater than one. 
It is thus desirable to study weakly interacting systems with exotic 
excitations if there is a chance that they can be realized in a laboratory. 
Our proposal for such a system is depicted in Fig.\ \ref{fig1}(a). It consists 
of a type-II superconducting film grown on top of a semiconductor 
heterostructure (such as GaAs/AlGaAs)
hosting a 2D electron gas. When a magnetic field is applied perpendicular 
to this planar device, an Abrikosov lattice of vortices forms in the 
superconductor. This results in a periodic modulation of the magnetic
field in the 2DEG with one-half of a magnetic flux quantum $\Phi_0=hc/e$ per unit 
cell. Devices like this have in fact been fabricated previously 
\cite{klitzing1,geim1,rojo1}
and studied theoretically \cite{rammer1,berciu1}.

A key new aspect of our proposal is a periodic array of pinning sites
imprinted on the superconductor. Pinning sites are regions of weakened
superconductivity, which attract vortex cores and can be created in a variety
of ways \cite{welp1}. When the number of vortices in the sample equals the 
number
of pinning sites, each pin is occupied by a vortex and a periodic
vortex lattice arises. The Magnetic field strength at which this happens is called the 
matching
field $B_M$. When the field is slightly below (above) $B_M$, vacancies
(interstitials) appear in the vortex lattice as illustrated in 
Fig. \ref{fig1}(b). 
Such defects in the vortex lattice then lead to a localized
deficit or surplus of magnetic flux quantized in the units of $\Phi_0/2$
shown in Fig. \ref{fig1}(c).  

Our principal claim, which we justify below by a general argument and
a detailed computation, is this: When the lowest Landau level 
of the 2DEG is filled by spin-polarized electrons, 
i.e. at filling fraction $\nu=1$, a defect in the
vortex lattice binds fractional electric charge in 
the 2DEG with the exact value $-e/2$ for a vacancy and $+e/2$ for an interstitial. 
Such a defect constitutes a literal realization of Wilczek's model of an anyon as
a bound state of magnetic flux and electric charge \cite{wilczek1}.
The exchange phase of this anyon $\pi/4$.

The flux deficit/surplus can be, at least in principle,
created, detected, and manipulated by the suite of scanning SQUID and Hall bar
probes that have been developed over the years to 
image \cite{kirtley1,tafuri1,kam1} and manipulate \cite{kam2,kam3} 
vortices in superconducting films and crystals. The proposed device could thus 
offer a unique opportunity to manipulate anyons with an unprecedented degree 
of control and advance the quest for a topologically protected quantum computer.

In order to execute a useful quantum computation, non-Abelian anyons are needed
\cite{kitaev1}. We remark that the device pictured in Fig.\ \ref{fig1}
could prove useful for the manipulation of such non-Abelian anyons if the
2DEG were put into the $\nu={5\over 2}$ FQH state, which is believed to be
described by the Moore-Read ``Pfaffian'' state \cite{moore1,nayak1,read1}.
In that situation, vacancies and interstitials should bind non-Abelian 
anyons and the scanning SQUID probe could be used to locate these and 
possibly perform the braiding operations required to implement a fault
tolerant quantum algorithm.

\section{Anyons bound to vacancies and interstitials: general arguments}

So why does a vacancy bind fractional charge? The answer can be given at 
several levels of sophistication and is essentially implicit in the body 
of work on the FQH effect. The simplest way to see that this is true is to recall
that electrons in the quantum Hall state at $\nu=1$ form an 
{\em incompressible fluid}. 
This means that in the ground state the electron charge 
density $\rho(\br)$ tracks 
the magnetic field strength $B(\br)$ in accordance with the defining relation
\begin{equation}\label{nu1}
\nu={N_e\over N_\Phi}={\rho\Phi_0\over eB},
\end{equation}
where $N_e$ and $N_\Phi$ denote the total number of electrons and flux quanta
in the sample, respectively. In an incompressible state the above relation
remains valid, to a good approximation, even for a spatially varying magnetic
field, as long as the variation is weak. In our device, a slowly varying field
occurs in the limit when the magnetic penetration depth $\lambda\gg a$, $a$ being 
the intervortex distance. In this limit, one obtains 
\begin{equation}\label{nu2}
\rho(\br)\simeq \nu e{B(\br)\over \Phi_0}.
\end{equation}
This relation, which becomes exact when integrated over the region containing the vacancy, implies
that at $\nu=1$ a deficit of one half flux quantum
produces a deficit of one half of an electron in the 2DEG. 

To see more clearly how the charge deficit/surplus
comes about we may employ a version
of Laughlin's argument \cite{laughlin3} similar to that used by Halperin
to deduce the existence of edge states in integer quantum Hall liquids 
\cite{halperin1}. The idea is to start from a
perfect vortex lattice and imagine creating an interstitial by adiabatically 
tuning the flux of the extra vortex from zero to $\Phi_0/2$.
We again consider the limit $\lambda\gg a$.

According to Faraday's law, $\nabla\times\bE=-(1/c)
(\partial \bB/\partial t)$, the time dependent magnetic field associated
with introducing the extra vortex induces an electric field with concentric
field lines as shown in Fig.\ \ref{fig1}(d). In the quantum Hall liquid such an 
electric field produces current strictly perpendicular to the field, 
\begin{equation}\label{hall}
\bj=\sigma_{xy}(\bE\times\hat{z}),
\end{equation}
where $\sigma_{xy}=\nu e^2/h$ is the
quantized Hall conductance. Using the continuity equation for the electric 
charge, one can integrate this current to find the rate of change
of the total charge $Q$ in the region bounded by an arbitrary closed 
contour ${\cal C}$ enclosing the flux,
\begin{equation}\label{dq1}
{dQ\over dt}= -\oint_{\cal C} dl \ {\bf n}\cdot\bj = -\sigma_{xy}\oint_{\cal C}
d{\bf l}\cdot\bE,
\end{equation}
where ${\bf n}$ is a unit vector normal to the contour. Using Stokes theorem, 
the last integral can then be writen as an integral of $\nabla\times\bE $ 
over an area bounded by ${\cal C}$. This we can evaluate 
with help of Faraday's law to obtain $dQ/dt=(\sigma_{xy}/c)d\Phi/dt$. 
It follows that the charge $\delta Q$ accumulated inside 
${\cal C}$ during the process of adiabatic flux insertion is
\begin{equation}\label{dq2}
\delta Q=e\nu {\delta\Phi\over \Phi_0}.
\end{equation}
For $\nu=1$ and $\delta\Phi=\pm\Phi_0/2$ the accumulated charge is $\pm e/2$.
It is to be noted that this argument depends only on the fundamental property
(\ref{hall})  and the above result (\ref{dq2}) should thus be valid
unconditionally as long as the 2DEG remains in the quantum Hall plateau.

Vacancies and interstitials, being bound states of charge and flux, 
$(-e/2, -\Phi_0/2)$ and  $(e/2, \Phi_0/2)$ respectively, acquire 
nontrivial Aharonov-Bohm phases upon adiabatic exchange, and can thus be thought of
as anyons. The standard textbook
counting procedure \cite{wilczek2}, according to which a charge $q$ taken around 
a flux $\Phi$ acquires a phase $2\pi(q/e)(\Phi/\Phi_0)$, would imply a ``semionic'' 
exchange phase 
$\theta=\pi/2$ for both vacancies and interstitials, and a mutual exchange phase 
$\theta_M=-\pi/2$. This conclusion, however, is incorrect for the following
reason. The charge $\pm e/2$ is built up from fermionic degrees of freedom and
thus has attached to it half of an electron spin, or more precisely, half
of the electron statistical phase, which needs to be taken into account when 
computing the exchange phase of the composite object. The simplest way to 
arrive at the correct result is to consider fusion rules for anyons. The
relevant rule \cite{wilczek2} states that the exchange phase $\Theta$ of a 
particle formed by combining $n$ identical anyons with exchange phase $\theta$ is 
$\Theta=n^2\theta$. 

Consider bringing together two interstitials. The
resulting object is $(e,\Phi_0)$, a bound state of an electron and a full flux 
quantum. The exchange phase of this object is $\pi+2\pi=3\pi$, the first term
reflecting the intrinsic electron exchange phase and the second the 
Aharonov-Bohm phase. But a $3\pi$ phase is equivalent to $\pi$, i.e.\ the bound 
state is a fermion. The above fusion rule with $n=2$ thus determines the 
exchange phase of an interstitial to be $\theta=\pi/4$. A similar argument
leads to the identical result for a vacancy. These results will be confirmed 
below by explicit model calculations.


\section{Continuum model}

\begin{figure*}
\includegraphics[width = 17.0cm]{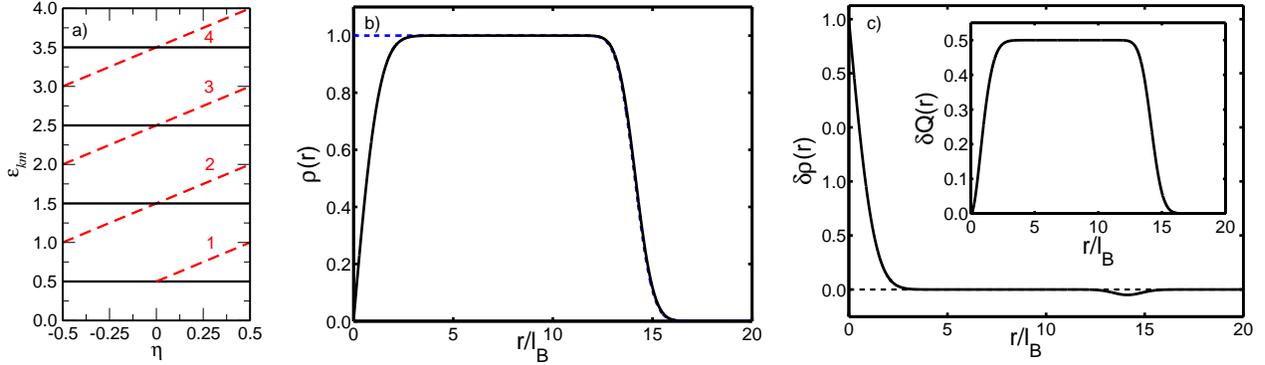}
\caption{(a) The energy levels (in units of $\hbar\omega_c$) of Hamiltonian
(\ref{h1}) as a function 
of fractional flux $\eta$. The numerals denote degeneracy of bound state
levels represented by dashed lines. Degeneracy of the extended states (solid
lines) is proportional to the area of the system. 
(b) Charge density $\rho(r)$, in units of 
$e/2\pi\ell_B^2$,  for $N=100$ as a function of distance $r$ from the 
origin. Solid line represents density $\rho_0(r)$ for uniform magnetic field, 
dashed line has a flux tube carrying $-\Phi_0/2$ at the origin. (c)
The excess charge density $\delta\rho(r)=\rho_0(r)-\rho(r)$. Inset shows 
the accumulated excess charge $\delta Q(r)=2\pi\int_0^r r'dr'\delta\rho(r')$
in units of $e$.  
}
\label{fig2}
\end{figure*}
We first consider a very simple continuum model, which is exactly soluble
and captures the essence of the effect discussed above. Consider the
following 2-dimensional electron Hamiltonian,  
\begin{equation}\label{h1}
{\cal H}={1\over 2m_e} \left(\bp-{e\over c}\bA\right)^2,
\end{equation}
with $m_e$ the electron mass, $\bp$ the momentum operator in the 
$x$-$y$ plane and $\bA=\bA_0+\delta\bA$. Here 
$\bA_0={1\over 2}B_0(\br\times \hat{z})$ represents a uniform field in the 
$\hat z$ direction and 
\begin{equation}\label{dA}
\delta\bA={\eta\Phi_0\over 2\pi r^2}(\br\times \hat{z}).
\end{equation}
The total field seen by an electron is then
\[
\bB(\br)=\nabla\times\bA=
\hat{z}B_0+\hat{z}\eta\Phi_0\delta(\br).
\]
The $\delta$-function 
serves as a crude representation of the flux added by an interstitial 
$(\eta={1\over 2})$ or removed by a vacancy $(\eta=-{1\over 2})$
 in the vortex lattice. 

It is straightforward to find the single particle eigenstates $\psi_{km}(\br)$
of ${\cal H}$ for arbitrary real $\eta$ by working  in the polar coordinate 
basis. They are labeled by the
principal quantum number $k=0,1,2,\dots$ and an integer angular momentum $m$. 
The allowed energy levels read
\begin{equation}\label{e1}
\epsilon_{km}={1\over 2}\hbar\omega_c\left[ 2k+1+|m-\eta|-(m-\eta)\right],
\end{equation}
where $\omega_c=eB_0/m_ec$ is the cyclotron frequency. 
The spectrum (\ref{e1}) is depicted in Fig.\ \ref{fig2}(a). For non-integer
$\eta$, in addition to the usual Landau levels, there also exist bound states 
associated with the extra flux. 

The eigenstates $\psi_{0m}$ in the lowest Landau level have a simple form
\begin{equation}\label{psi0}
\psi_{0m}(z)=A_m|z|^{-\eta} z^m e^{-|z|^2/4},
\end{equation}
where $z=(x+iy)/\ell_B$ is the dimensionless complex coordinate, 
$A_m=[\ell_B^22\pi 2^{m-\eta}\Gamma(1+m-\eta)]^{-1/2}$ is the normalization 
constant and $\ell_B=\sqrt{\hbar c/e B}$ the magnetic length. 
If we fill the lowest 
Landau level with electrons, then the many-body wavefunction can be constructed
as a Slater determinant of $\psi_{0m}(z_i)$, where $z_i$ is a complex 
coordinate of the $i$-th electron. The usual simplifications using the
properties of Vandermonde determinants then give
\begin{equation}\label{Psi0}
\Psi(\{z_i\})={\cal N}\prod_i|z_i|^{-\eta}\prod_{i<j} (z_i-z_j) e^{-\sum_i|z_i|^2/4}.
\end{equation}

We are interested in the charge density implied by the above many-body 
wavefunction, 
which is given as the expectation value $\rho(\br)=\langle\Psi_0|\hat{\rho}
|\Psi_0\rangle$ of the electron density operator 
$\hat{\rho}=e\sum_i\delta(\br-\br_i)$. For a droplet composed of $N$ electrons 
occupying the $N$ lowest angular momentum states we obtain 
\begin{equation}\label{rho1}
\rho(\br)=e\sum_{m=0}^{N-1}|\psi_{0m}(\br)|^2.
\end{equation}
The charge density obtained by numerically evaluating Eq.\ (\ref{rho1})
is displayed in Fig.\ \ref{fig2}, both for a uniform field $(\eta=0)$ and the 
vacancy $(\eta=-{1\over 2})$ placed at the origin. 
We observe that the inclusion of a vacancy 
leads to suppression of the charge density near the origin, the excess charge 
being deposited at the outer edge of the system. The integrated 
charge deficit $\delta Q(r)$
quickly saturates at $e/2$, confirming the conclusion reached above by 
general argument. The vacancy in this model indeed binds half of an electron.

From the many-body wavefunction (\ref{Psi0}) one can obtain the charge 
and the statistical angle associated with a vacancy by the standard calculation
\cite{arovas1} of the Berry phase.  To compute the Berry phase it is 
convenient to perform a gauge transformation
\begin{equation}\label{gauge}
\delta\bA\to\delta\bA-{\eta\Phi_0\over 2\pi}\nabla\varphi
\end{equation}
where $z=|z|e^{i\varphi}$. The resulting vector potential can be seen to vanish
everywhere, except for a cut along the positive $x$ axis. The effect of the 
gauge transformation is to multiply
the eigenstates (\ref{psi0}) by $e^{-i\eta\varphi}$ which has the effect of 
replacing $|z|^{-\eta}$ by $z^{-\eta}$. In this ``string gauge'' the entire phase
information is encoded in the wavefunction. The many-body wavefunction
for a vacancy $(\nu=-{1\over 2})$ located at the complex coordinate $w$ can 
thus be written as  
\begin{equation}\label{Psiw}
\Psi_w(\{z_i\})={\cal N}_w\prod_i(z_i-w)^{1/2}\Psi_0(\{z_i\}),
\end{equation}
where $\Psi_0(\{z_i\})=\prod_{i<j} (z_i-z_j) e^{-\sum_i|z_i|^2/4}$.

If we now adiabatically carry the vacancy along a closed contour ${\cal C}$ 
the wavefunction acquires a purely geometric phase
\begin{equation}\label{Berry1}
\gamma({\cal C})=i\oint_{\cal C}dw\langle\Psi_w|{\partial\over\partial w}
\Psi_w\rangle. 
\end{equation}
Calculation of $\gamma({\cal C})$ proceeds exactly as in Ref.\ \cite{arovas1}
and yields
\begin{equation}\label{Berry2}
\gamma({\cal C})=-\pi{\Phi\over \Phi_0},
\end{equation}
where $\Phi$ is the total magnetic flux enclosed by the contour ${\cal C}$. 
This result coincides with the Aharonov-Bohm phase acquired by a charge 
$-e/2$ carried along a trajectory enclosing flux $\Phi$, confirming once again
our earlier result for the electric charge of the vacancy.

For two vacancies, located at $w_a$ and $w_b$ the many-body wavefunction
reads
\begin{equation}\label{Psiwab}
\Psi_{w_aw_b}={\cal N}_{w_aw_b}\prod_i(z_i-w_a)^{1/2}(z_i-w_b)^{1/2}\Psi_0.
\end{equation}
If we now carry $w_a$ along a contour ${\cal C}$ that encloses $w_b$, the
latter held static, then Eq.\ (\ref{Berry1}) gives the Berry phase
\begin{equation}\label{Berry3}
\gamma({\cal C})=-\pi\left({\Phi\over \Phi_0}-{1\over 2}\right).
\end{equation}
The first term reflects the Aharonov-Bohm phase as before. Since the encircling 
operation can be thought of as two consecutive exchanges we must interpret the 
second term as twice the exchange phase of the vacancy, $\theta=\pi/4$.


\section{Lattice model}

Thus far we have ignored the periodic variation of the magnetic field
induced by the vortex lattice. A question thus arises whether this could 
affect the emergence of anyons in realistic systems. To see that this is not 
the case we now consider a model which describes the limit of
strong field variation and also includes the effect of a Zeeman interaction.
\begin{figure*}
\includegraphics[width = 16.8cm]{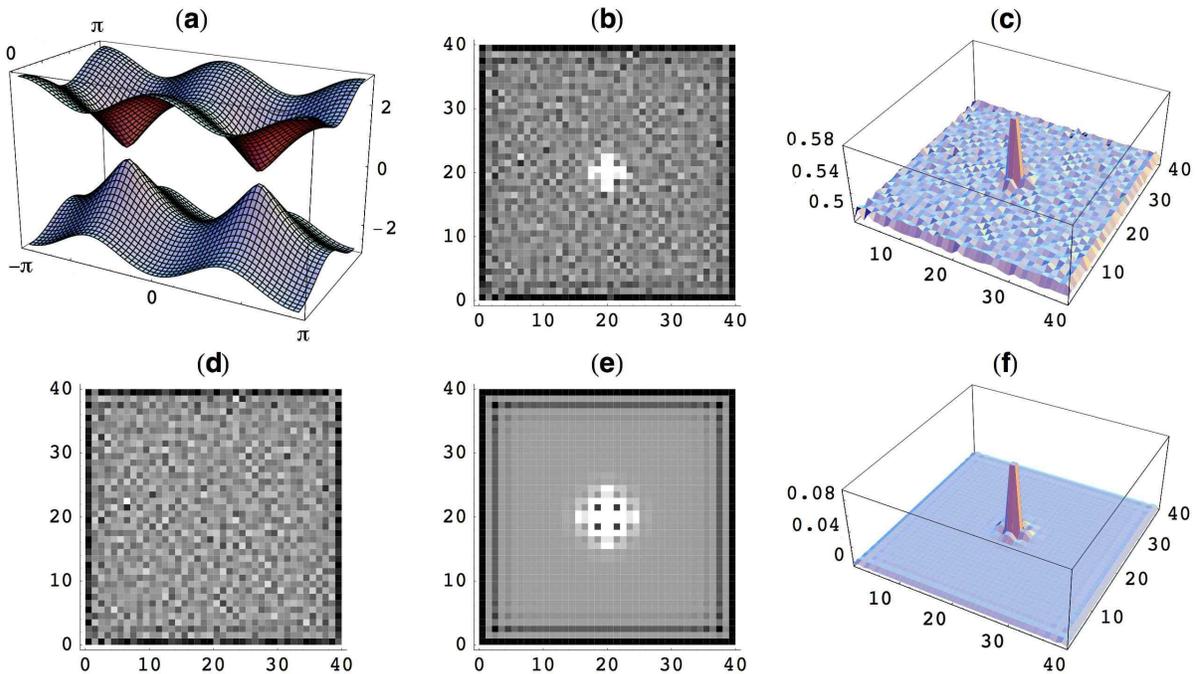}
\caption{a) Band structure of the tight binding model on a square lattice 
with one half flux quantum threading each plaquette and $t'=0.05t$. b,c) 
Charge density 
$\rho$ at half filling with random on-site energies (Gaussian distribution 
with $\Delta\epsilon=0.05t'$) and a full flux quantum modeling the interstitial 
piercing the central plaquette. d) Charge density $\rho_0$ with the same 
realization of randomness but without the extra flux. e,f) The charge density
$\delta\rho=\rho-\rho_0$ induced by the interstitial. The induced charge
$\delta Q$ integrates to $e/2$ to within machine accuracy.
}
\label{fig4}
\end{figure*}

If the effective Zeeman coupling in a 2DEG is sufficiently strong, then,
in addition to a quasi-periodic vector potential, the electrons also feel a 
periodic 
scalar potential. This effect becomes important in diluted magnetic
semiconductors, such as Ga$_{1-x}$Mn$_x$As or Cd$_{1-x}$Mn$_x$Te, where the 
effective gyro-magnetic ratio can be of the order $\sim 10^2-10^3$. As argued 
in Ref. \cite{berciu1}, in the limit $\lambda\ll a$, electrons in a 2DEG 
become almost localized near the maxima
of the Zeeman field and can be described by a
tight binding model with one-half magnetic flux quantum per unit cell. 

This motivates our consideration of a tight binding Hamiltonian
\begin{equation}\label{ht1}
{\cal H}=-\sum_{ij}(t_{ij}e^{i\theta_{ij}}c_j^\dagger c_i +{\rm h.c.})
+\sum_i\epsilon_ic_i^\dagger c_i,
\end{equation}
where $c_i^\dagger$ creates an electron on a site $\br_i$ of a square lattice,
and $t_{ij}$ and $\epsilon_i$ are real hopping amplitudes and on-site energies, 
respectively. The effect of magnetic field is included by the Peierls phase
factors
\begin{equation}\label{th}
\theta_{ij}={2\pi\over\Phi_0}\int_{\br_i}^{\br_j}\bA\cdot{\bf dl}.
\end{equation}
In the following we consider the case of nearest and next-nearest
neighbor hopping amplitudes $t$ and $t'$ respectively, but note the addition of 
longer range hoppings does not affect our results.

In the absence of defects we have $\epsilon_i=\epsilon=$const and the
magnetic flux through each plaquette is $\Phi_0/2$. 
The spectrum of the Hamiltonian  (\ref{ht1}) 
consists of two branches, reflecting the fact that the unit cell now contains 
two lattice sites. In the Landau gauge, $\bA=B_0\hat{y}x$ with 
$B_0=\Phi_0/2a^2$, the allowed energies read
\begin{equation}\label{ek}
E_\bk=\epsilon\pm 2t\left[\cos^2{k_x}+\cos^2k_y+4\gamma^2\sin^2k_x\sin^2k_y\right]^{1/2},
\end{equation}
where  $\bk$ is a wavevector drawn from the first magnetic 
Brillouin zone, $\gamma=t'/t$,  and we have set
$a=1$. For $t'=0$ the spectrum exhibits two Dirac points at $(\pm\pi/2,\pi/2)$.
Inclusion of next-nearest neighbor hopping breaks the time reversal symmetry
and produces a gap at these two points, as illustrated in Fig.\ \ref{fig4}(a). 
We are interested in the half-filled case
with the lower band filled and the upper band empty. This 
situation is analogous to the $\nu=1$  continuum case with 
Hall conductance $\sigma_{xy}=e^2/h$ discussed previously. 

A vacancy in the vortex lattice removes flux $\Phi_0/2$ from the 
plaquettes adjacent to the affected site,
modifies its on-site energy $\epsilon_i$ and possibly also those of the nearby
sites. While
the total flux removed is exactly $\Phi_0/2$, the change of the on-site 
energies $\delta\epsilon_i$
depends on the details of the underlying microscopic model for the Zeeman 
interaction. We find that, as expected by the general argument, the total 
charge deficit associated
with a vacancy depends critically on the total flux removed but is
independent of $\delta\epsilon_i$ or the details of the flux distribution. 

We have solved the model specified by Hamiltonian (\ref{ht1}) by exact 
numerical diagonalization for lattice sizes up to $50\times 50$ sites. 
We have considered various configurations of $\delta\epsilon_i$. In all cases 
we found that the total charge induced by an interstitial/vacancy is $\pm e/2$,
to within machine accuracy. This result continues to hold even in the
presence of {\em randomness} in the on-site energies, as long as the randomness
is weak compared to the gap. This is illustrated in Fig.\ \ref{fig4}(b-f).

\section{Summary and conclusions}

Our foregoing discussion shows how two simple and
well understood systems, a type-II superconductor and a 
2-dimensional electron gas in the integer quantum Hall regime, can 
exhibit some rather unusual properties when brought into close 
proximity. These include nontrivial exchange statistics and  charge 
fractionalization, properties that are normally thought of as hallmarks
of strong correlation physics. In our proposed device these phenomena occur
in what must be characterized as a weakly correlated system: the many-body 
wavefunctions of electrons in both subsystems 
are conventional Slater determinants of single-particle states. 

There are two key ingredients that give rise to the above unusual 
properties: (i) precise flux quantization in a type-II superconductor and,
(ii) incompressibility of the 2DEG at integer filling $\nu$. The device 
displayed in Fig.\ \ref{fig1} is designed to project a precisely
quantized deficit or surplus of magnetic flux onto the 2DEG, which 
responds by adjusting its many-body wavefunction to
produce a localized deficit or surplus of electric charge with precise
value $e/2$. This phenomenon can be regarded as a consequence of gauge 
invariance and is therefore robust against any weak but otherwise arbitrary 
perturbation imposed on the system. 

A question that we must ask is to what extent are the
fractional charges associated with the vacancies and interstitials 
legitimate quasiparticle excitations of the system. Surely, from the point of 
view of a 2DEG at integer filling, such objects are {\em not}  natural 
excitations. Indeed, in our device they occur in response to the local change 
in the magnetic flux, which is external to the 2DEG. The key point is that
fractional charges bound to defects are excitations
of the {\em entire system} in the same sense as fractional charges
bound to domain walls are excitations of a dimerized polyacetylene chain
\cite{su}. 

Fractional charges should be relatively easy to detect in a transport 
experiment in the regime where the longitudinal conductivity of the 2DEG is 
furnished solely by the $e/2$ defects. Fractional statistics, on the other
hand, will be more difficult to establish since this requires an interference
experiment; in essence we need a defect to quantum-delocalize and interfere
with itself after encircling another defect. As noted recently, non-trivial
statistics might be easier to detect in the $\nu={5\over 2}$ non-Abelian
case \cite{nayak3,kitaev3}.

Aside from demonstrating the possibility of unconventional phenomena in a weakly
interacting system, our proposed device could allow for the manipulation of anyons
in the context of quantum information processing. Indeed, some schemes for 
fault tolerant topological quantum computation \cite{kitaev1,freedman1,kitaev2}
involve execution of braiding and fusion operations on anyons. In the
present setup these could  be facilitated by the fact that our anyons are 
permanently bound to physical magnetic fluxes and well developed 
techniques exist to detect and manipulate these. As already remarked above if 
the 2DEG in our device were tuned into the $\nu={5\over 2}$ Moore-Read Pfaffian 
state then a defect in the vortex lattice would bind a quasiparticle with
non-Abelian exchange statistics. Having a ``handle'' on such a quasiparticle
in the form of the attached magnetic flux could prove useful for realizing 
the requisite braiding and fusion operations.

\smallskip
\noindent
{\it Acknowledgments\/} ---  
The authors are indebted to M. Berciu and A. Vishwanath for some key suggestions
and wish to thank I. Affleck, D. Fisher, D. Haldane, C. Kallin, A. Kitaev, 
K. Shtengel, D. Scalapino and Z. Tesanovic for 
stimulating discussions and correspondence. The work reported here was
supported by NSERC, CIAR and the Killam Foundation.



\end{document}